\documentclass[
reprint,                     
superscriptaddress,
 amsmath,amssymb,
 aps,
prl,
]{revtex4-1}

\usepackage{graphicx} 
\usepackage{dcolumn}  
\usepackage{bm}       
\usepackage{xcolor}

\begin{document}


\title{High-frequency nonlinear conductivity of a Wigner crystal}

\author{A.J.~Schleusner}
\email{schleusn@msu.edu}
\affiliation{
 Department of Physics and Astronomy, Michigan State University, East Lansing, MI 48824, USA}
\author{M.T.~Elewa}
\affiliation{
 Department of Physics and Astronomy, Michigan State University, East Lansing, MI 48824, USA}
\author{N.R.~Beysengulov}
\affiliation{
 Department of Physics and Astronomy, Michigan State University, East Lansing, MI 48824, USA}
\affiliation{
 EeroQ Corporation, Chicago, IL 60651, USA}
\author{C.A.~Mikolas}
\affiliation{
 Department of Physics and Astronomy, Michigan State University, East Lansing, MI 48824, USA}
\author{J.~Pollanen}
\email{pollanen@msu.edu}
\affiliation{
 Department of Physics and Astronomy, Michigan State University, East Lansing, MI 48824, USA}

\date{\today}

\begin{abstract}
Electrons trapped above the surface of superfluid helium are a disorder-free platform for investigating the formation and dynamics of low-dimensional Wigner crystals. A characteristic nonlinear transport feature of this electronic solid suspended above the helium surface is the Bragg-Cherenkov effect, in which the mobility of the smoothly moving crystal is limited by the coherent emission of helium surface waves (ripplons). The effect has been understood in the conventional Cherenkov setting in which the crystal moves at a constant speed. 
Here we report on transport measurements of electrons on helium confined in a microchannel geometry to investigate the non-equilibrium response of the Wigner solid when it is subjected to a high-frequency driving field. 
Surprisingly, the experiments reveal a strongly nonlinear transport response of the confined Wigner solid at frequencies nearly an order of magnitude larger than the ripplon frequencies contributing to the conventional Bragg-Cherenkov effect. We relate this observation to the coupling of the Wigner solid to ripplons with higher-order Bragg vectors, which gives rise to a dynamical friction that provides a mechanism for the observed high-frequency pinning.
\end{abstract}

\maketitle

When the repulsive Coulomb interaction between electrons exceeds their kinetic energy, they will self-organize into an ordered state known as the Wigner solid~\cite{Wigner34}. The fundamental nature of this electron solid state results from the subtle interplay between interactions, dimensionality, spatial confinement, and disorder. Evidence for the formation of Wigner solids has been reported in a wide variety of low-dimensional systems, including semiconductor heterostructures~\cite{Goldman90,Piot08,Falson22,Shayegan22}, graphene-based devices~\cite{Kumar18,Tsui24}, layered transition metal dichalcogenides~\cite{Smoleński21,Zhou21,Xiang24arx,Wang2025spectroscopywignercrystalpolarons,Zhang2025wignerpolaronsrevealwigner}, and Moir{\'e} structures~\cite{Regan20,Li21}. Among the diverse material systems purportedly exhibiting Wigner crystals, electrons trapped above the surface of superfluid helium host the most pristine examples of two-dimensional~\cite{Grimes79} and quasi-one-dimensional electron solids~\cite{Rees16B,Rees17}. 

In the low-dimensional system of electrons on helium, the Coulomb interaction between electrons is largely unscreened due to the isolation provided by the superfluid substrate. This produces a quantum non-degenerate system, the ground state of which is a Wigner solid~\cite{Monarkha2004}. Electrons floating above the helium surface experience an extremely low level of disorder, leading to the highest mobility in condensed matter~\cite{shirahama95a} and an electronic solid that is unpinned by residual defects. Below $\simeq0.7\text{~K}$, the scattering from helium vapor atoms is negligible and the mobility of the electron system is limited by its interaction with the quantum field of helium capillary waves, i.e. ripplons. In this regime, the Wigner solid on the surface of helium exhibits a strange and highly nonlinear transport response~\cite{Giannetta91,Shirahama95b,Kristensen96}. As the electron system is driven by a constant external force, its velocity saturates once it reaches the velocity of ripplons having a wavelength matching the inter-electron spacing. This phenomenon is known as the Bragg-Cherenkov effect and occurs when the Bragg lattice of the electron crystal Cherenkov radiates coherent ripplons that constructively interfere and apply a back-action force opposing the motion of the crystal~\cite{Dykman97}. While the electron system is non-degenerate, it is not purely classical. In many experiments, including those reported here, the temperature is below the corresponding energy of short wavelength phonons of the Wigner crystal. In this case, the conductivity is determined by quantum fluctuations, making it significantly different from the classical regime~\cite{Dykman1982}. Nevertheless, the conventional Bragg-Cherenkov mechanism still works for frequencies low compared to those of ripplons with wave vector equal to the reciprocal lattice vector, i.e. when the motion of the crystal can be consider quasi-constant. 

From the experimental side, open questions remain regarding the nature of the Wigner solid on helium and its interaction with the bosonic field of ripplons. For example, when subjected to a sufficiently large driving force, the electron solid transitions to a non-equilibrium state exceeding the Bragg-Cherenkov speed limit. This non-equilibrium state has been interpreted as resulting from either a dynamical melting of the crystal~\cite{Giannetta91,Syvokon08,Syvokon14} or a transient depinning~\cite{Vinen1999,Shirahama95b,Rees16L} of the Wigner solid from a commensurate lattice of dimples on the helium surface. Relatedly, it is not known how the electronic crystal and its coupling to the helium surface will behave when the driving frequency is high enough that the electron motion can no longer be treated as a quasistatic drift. To date, experiments investigating the Bragg-Cherenkov effect have used either a relatively low driving frequency~\cite{Shirahama04,Ikegami10,Syvokon10} or a smoothly increasing driving field~\cite{Rees16L,Rees17}. In contrast, here we report on experiments investigating the high-frequency transport response of a Wigner crystal far outside this quasistatic regime. Surprisingly, we find that the response of the Wigner crystal is independent of the drive amplitude even when the oscillating electron displacement is much smaller than the lattice spacing of the crystal. Additionally, in this high-frequency regime we observe a sharp transition with increasing drive strength to a highly conducting state consistent with an electronic liquid. To account for these observations we developed a theory of the nonlinear coupling between the Wigner crystal and the ripplon bath when the electrons are subjected to a rapidly oscillating driving field. In this regime the electrons experience a frictional force arising from the coupling of the electron lattice to higher frequency ripplon overtones.

\begin{figure}
    \centering
    \includegraphics[width = 8.5cm]{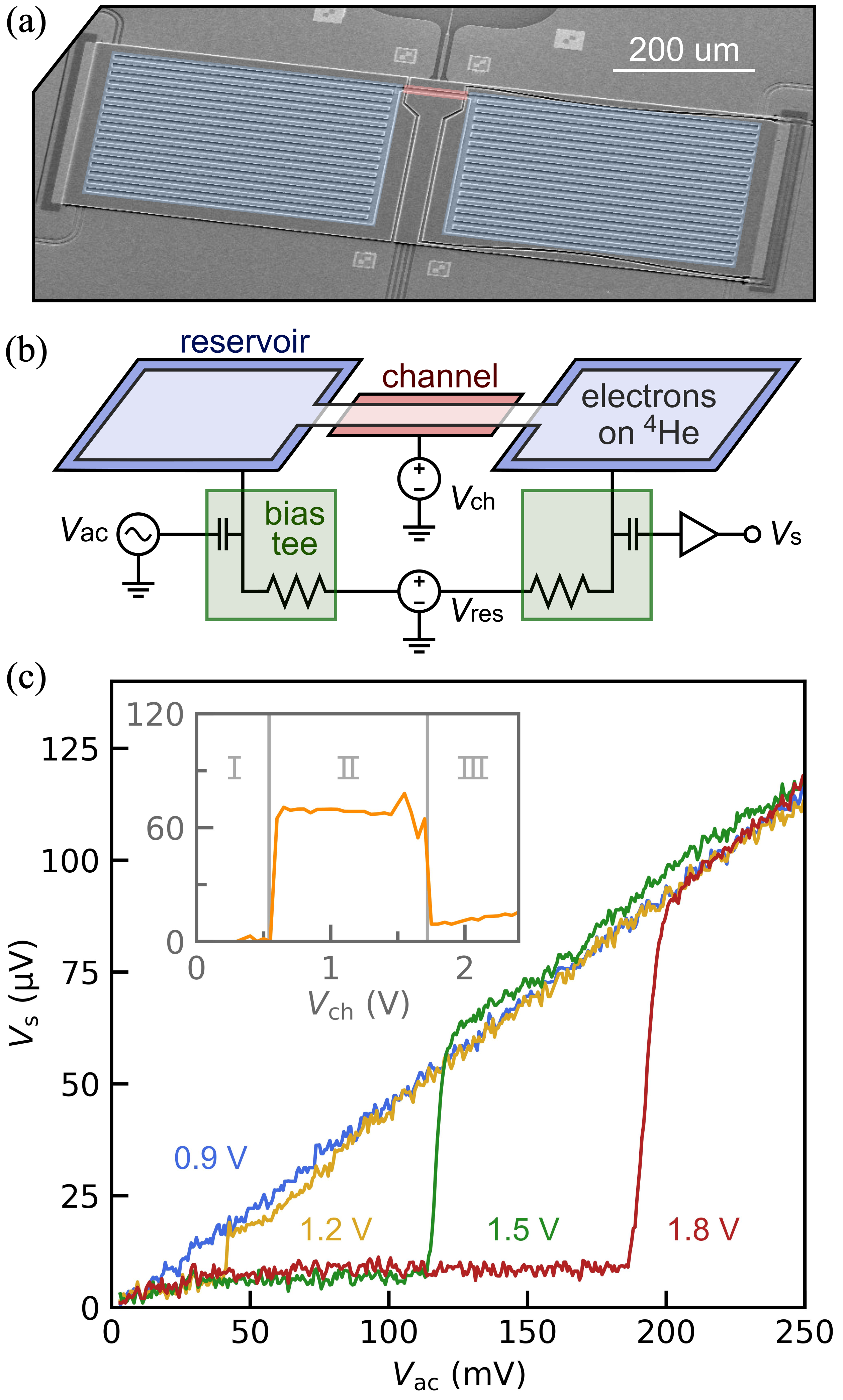}
    \caption{(a) A false-color scanning electron microscope image of the device showing the central electron confining microchannel (red) and electron reservoir (blue) regions. (b) Schematic of the electron system, electrodes, and voltages used to control the electron density and conduct transport measurements. (c)  Electron transport as a function of the driving field amplitude $V_{\textrm{ac}}$ at a frequency of 300~kHz for increasing central channel bias voltages ($V_{\textrm{ch}} =0.9, 1.2, 1.5, 1.8\ \text{V}$). 
    Inset: Transport signal as a function of $V_{\textrm{ch}}$ for $V_{\textrm{ac}} = 150\text{~mV}$. The transport regimes correspond to an empty channel (I), a high conductivity electron state (II), and a low-conductivity Wigner crystal (III). } 
    \label{fig1}
\end{figure}

To study the Wigner solid, we employ a microchannel device architecture~\cite{Rees16B,Mikolas25}, as shown in Fig.~\ref{fig1}a. The channeled regions of the device are filled with liquid helium via capillary action. Two large reservoir regions consist of many of these microfluidic channels in parallel and store hundreds of thousands to several million electrons. The density and spatial confinement of electrons in the central microchannel connecting the reservoirs is carefully controlled by tuning bias voltages applied to electrodes beneath the central channel~$V_\mathrm{ch}$ and the reservoirs~$V_\mathrm{res}$ (see Fig.~\ref{fig1}b). Furthermore, transport of electrons through the central microchannel is induced by capacitively coupling to an ac driving voltage~$V_{\text{ac}}$ applied to one of the two reservoir electrodes. The electron motion is detected as an induced voltage $V_{\text{s}}$ on the second reservoir electrode using phase-sensitive lock-in techniques. Since the central channel acts as a bottleneck for electron transport through the device, this measurement is particularly sensitive to the electronic response and structure within the central microchannel. 

The device used in the experiments was fabricated onto a 7~mm~$\times$~2~mm high-resistivity silicon chip.  Photolithography and thermal evaporation of Ti/Au were used to first create the central microchannel and reservoir electrodes. A subsequent layer of insulating resist was deposited and selectively etched to produce the 1.4~$\mu$m deep channels that comprise the electron reservoirs and the 90~$\mu$m long and 7~$\mu$m wide central microchannel (Fig.~\ref{fig1}a). The device was housed within a superfluid-leak-tight copper cell and attached to the mixing chamber stage of a dilution refrigerator. When cooled to $\simeq0.5~\text{K}$, helium was condensed into the sample cell via a capillary tube and electrons deposited onto the superfluid surface via thermal emission from a nearby tungsten filament. 

We first present transport measurements to establish the standard low-frequency response of the electron system and the conventional Bragg-Cherenkov effect. Fig.~\ref{fig1}c depicts the measurement signal for electron transport through the microchannel at a frequency of 300~kHz. By varying the central microchannel bias during this measurement, three distinct, and characteristic, transport regimes emerge as shown in the Fig.~\ref{fig1}c inset.  In regime I, the central microchannel potential exceeds the reservoir chemical potential, which results in no electrons occupying the central channel and consequently no transport signal, i.e.\ $V_{\text{s}}=0$. In regime II, the central microchannel electric potential is less than the reservoir chemical potential and the electron density in the channel increases as electrons flow in from the reservoirs. This regime exhibits a large transport response consistent with the formation of either a non-equilibrium electron liquid state~\cite{GIANNETTA1991,Mikolas25} or a dynamically unpinned Wigner crystal~\cite{Rees16L}. The transport signal in regime II is proportional to the amplitude of the driving field, as shown in the blue data trace taken at $V_{\textrm{ch}}=0.9$~V, indicative of a relatively weak interaction between the electrons and the helium surface. Increasing the electron density further ultimately results in a sharp decrease in $V_\text{s}$ once the electron system transitions into a low-conductivity Wigner solid (regime III). This is the regime of the conventional Bragg-Cherenkov effect, in which the electron velocity becomes independent of the driving field amplitude~\cite{Ikegami10,Ikegami12,Badrutdinov16}, as shown in the green and red data traces of Fig.~\ref{fig1}c. Here the speed of the crystal saturates at  $\approx 10$~m/s~\cite{Glasson01,Rees17}, the phase velocity of ripplons having a wavevector matching the magnitude ($G_1$) of the fundamental reciprocal lattice vector $\mathbf{G}_1$.  For the triangular crystal lattice, the magnitude of this vector is $G_1=2\pi\sqrt{2n_s/\sqrt{3}}$, where $n_s$ is the areal electron density determined from finite element modeling~\cite{Mikolas25}. For the typical electron density of the Wigner solid, the frequency of these ripplons is of order $\omega_{G{_1}}/2\pi \simeq 10$~MHz, determined from the ripplon dispersion relation $\omega_{G{_1}}=\sqrt{\sigma G_1^3/\rho}$, where $\sigma$ is the surface tension and $\rho$ is the density of helium~\cite{Fisher79}. At 300~kHz, well below $\omega_{G{_1}}/2\pi$, the motion of the crystal can be consider quasistatic as an electron will travel approximately $10~\mu\mathrm{m}$ per cycle of the driving field, or of order 100 lattice sites.

\begin{figure}
    \centering
    \includegraphics[width = 8.5cm]{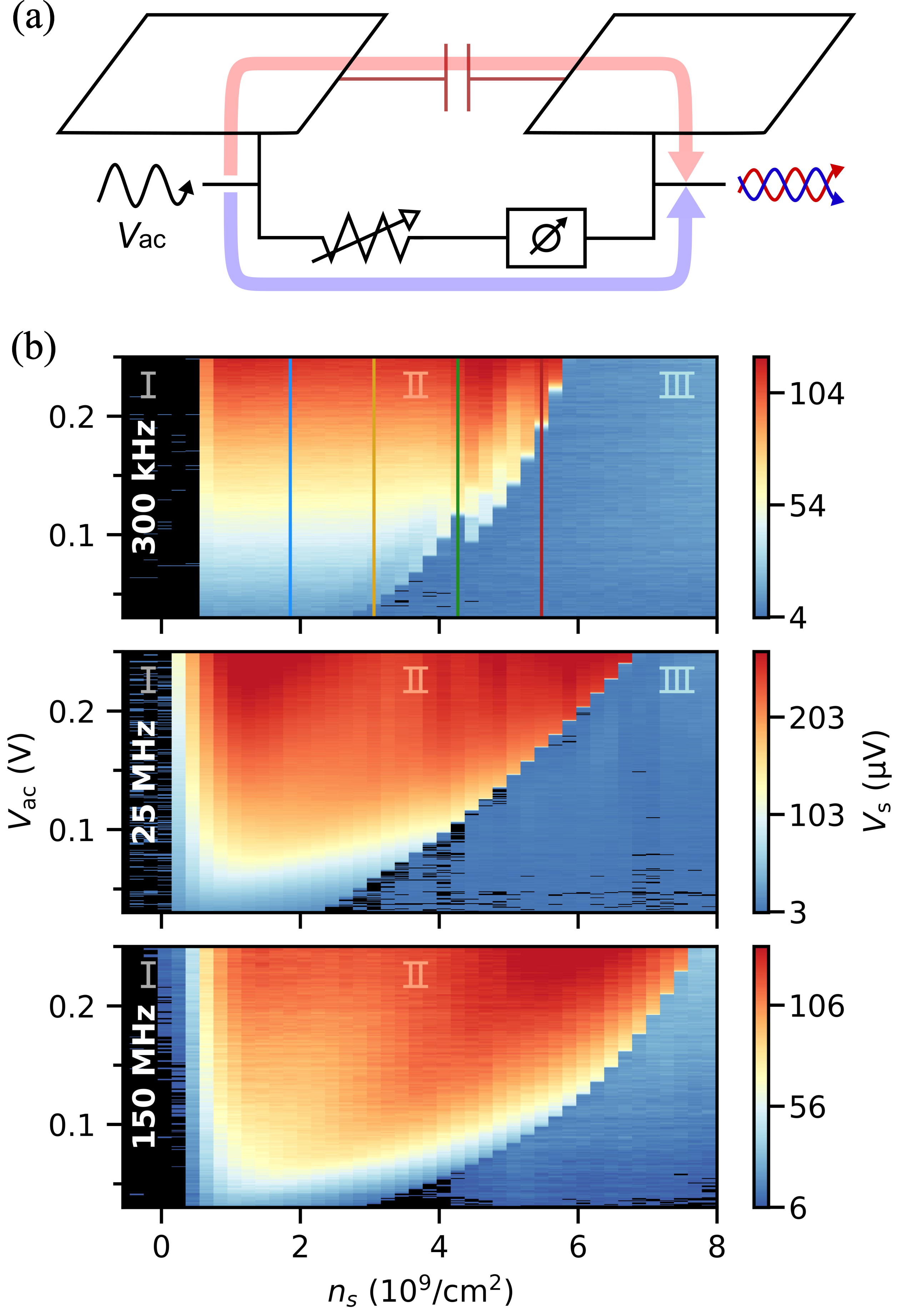}
    \caption{(a) Schematic of the compensation line approach employed to measure the high frequency transport of the microchannel-confined electron system. A voltage tunable attenuator and phase shifter are used to interferometrically reduce the parasitic crosstalk inherent to the device geometry. (b)  The induced transport voltage $V_\text{s}$ as a function of the central microchannel bias $V_\text{ch}$ and the ac driving voltage $V_\text{ac}$ for $f=0.3,25,150$~MHz driving fields.  The blue, gold, green, and red vertical lines in the top plot correspond to the line-cuts shown in Fig.~\ref{fig1}c.}
    \label{fig3}
\end{figure}

To escape this quasistatic regime requires transport measurements exceeding $\omega_{G{_1}}/2\pi$. Measuring the response of a Wigner crystal in a microchannel device at such frequencies is drastically hindered by the emergence of a parasitic background arising from the cross-talk between the driving and detection electrodes. To minimize this background, we incorporate a compensation line by placing a variable attenuator and phase shifter in parallel to the device. By tuning this compensation line when the central channel is free of electrons, we are able to produce a signal to interferometrically cancel this unwanted background (Fig.~\ref{fig3}a). This attenuates the background by approximately an order of magnitude (see Appendix~A) enabling the measurements described below.

Fig.~\ref{fig3}b shows the results of representative transport measurements taken while sweeping $V_{\text{ch}}$ and $V_{\text{ac}}$ at driving frequencies up to 150~MHz, well above the characteristic ripplon frequency. Interestingly, even at this high frequency, the electron system still exhibits a clear and sharp transition to a low conductivity Wigner crystal, characterized by a saturated velocity, as the density is increased. Similarly, starting from the Wigner crystalline state and increasing the drive at fixed density results in a sharp transition to a highly conducting state. This is despite the fact that the motion of the electron solid is far from quasistatic. At 150~MHz the typical displacement of an electron per cycle of the driving field is $\approx 30$~nm, which is $4-7$~times smaller than the lattice spacing of the crystal and the corresponding dimpled corrugation of the helium surface. This indicates that at high frequency the observed transition to the high-conductivity state as a function of drive amplitude is likely not associated with a decoupling of the electron lattice from the dimples. Rather, the data are consistent with a dynamical melting of the electron crystal as the amplitude of the drive is increased. Based on this interpretation, we can estimate the effective electron temperature along the critical density transition between regime II and III from the ratio of the electron–electron interaction energy and the average kinetic energy $\Gamma =E_\mathrm{C}/k_\mathrm{B}T$. For a two-dimensional system of electrons on helium the transition between the electron liquid and crystal occurs at a critical value of $\Gamma \approx 130$~\cite{Monarkha2004}, which leads to an effective melting temperature $T_{\textrm{eff}}=e^2\sqrt{\pi n_s}/(4\pi\epsilon_0\cdot130~k_\mathrm{B}) \simeq 1-2$~K for the 150~MHz data in Fig.~\ref{fig3}b.

\begin{figure}
    \centering
    \includegraphics[width = 8.5cm]{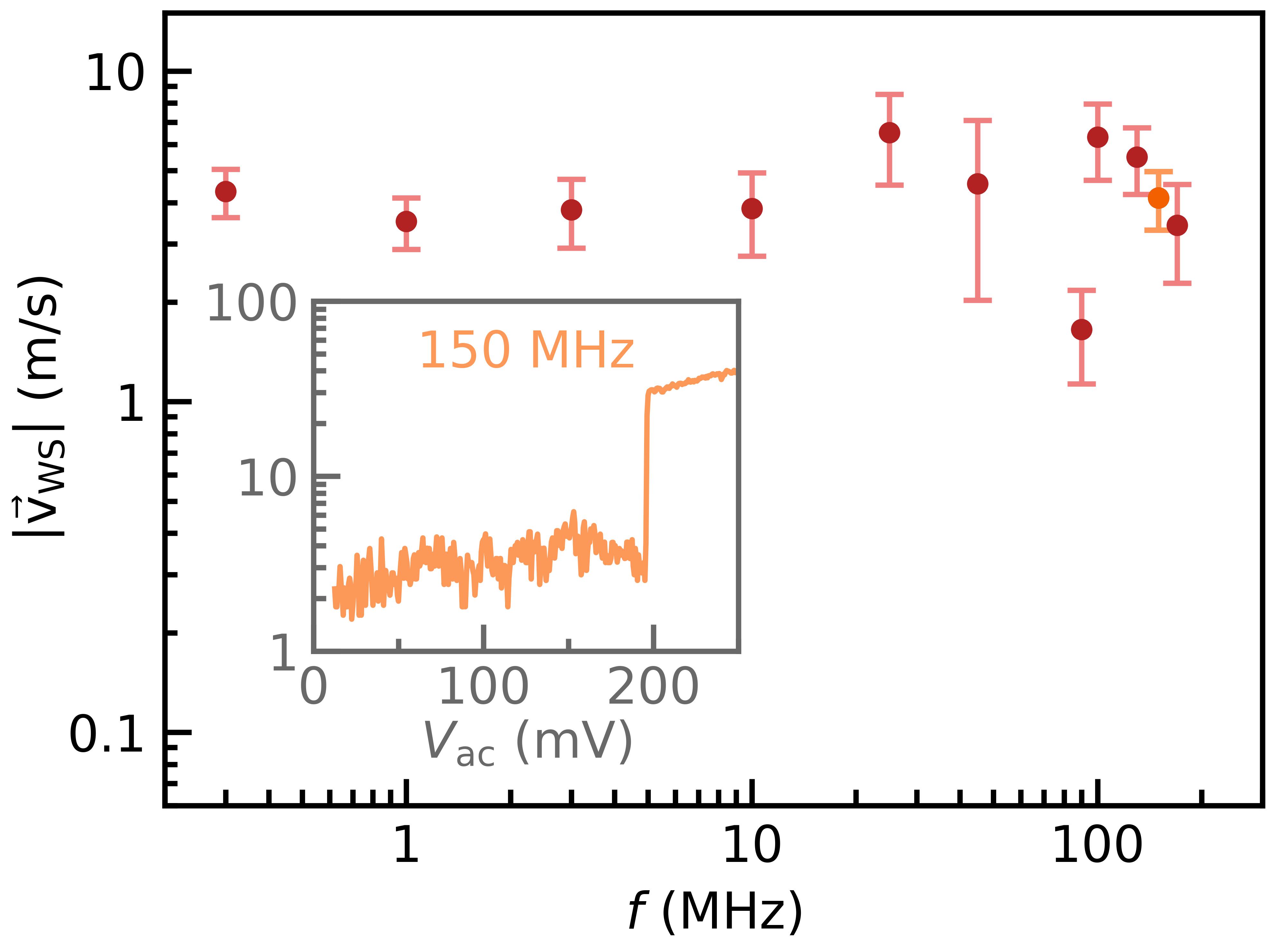}
    \caption{Velocity of a Wigner solid moving along the central microchannel. For these data, the electron density in the channel is $7.1\times10^9/\text{cm}^2$. Inset: Velocity of the Wigner solid as a function of the driving voltage amplitude at a frequency of 150~MHz. The data trace shows the saturated velocity response when the electrons in the microchannel are crystallized as well as a sharp transition to a non-equilibrium state of the electron system that moves at a higher velocity.}
    \label{fig4}
\end{figure}

 We further confirm that the saturation of the Wigner crystal motion at high-frequency results from its interaction with the helium surface by estimating the speed with which the electron crystal moves. For an electron lattice spacing $d$ the phase velocity of the ripplons contributing to the Bragg-Cherenkov effect is $\left|\vec{v}_{G{_1}}\right|=\sqrt{2\pi\sigma/d\rho}$. For our measurements this yields a phase velocity of order several m/s. Separately, we can calculate the velocity of the Wigner solid from the measured voltage plateau using a standard lumped element approximation of the microchannel device~\cite{Lin18}. As shown in Fig.~\ref{fig4} we find the velocity of the Wigner solid $\left|\vec{v}_{\textrm{WS}}\right|$ to be of order $1-10$~m/s, consistent with the propagation speed of helium surface excitations. 
 


To understand these experimental observations, we extend the Bragg-Cherenkov theory of Ref.~\cite{Dykman97} to the case of a Wigner solid driven by a high-frequency electric field $\mathbf{E}(t)=\mathbf{E}_0\sin(\omega t)$ that causes the electrons to oscillate with displacement amplitude $\mathbf a=e\mathbf E_0/m\omega^2$. The electron-ripplon interaction Hamiltonian takes the form $H_i = \sum_{\mathbf{q}}V_{\mathbf q}\rho_{\mathbf q} (b_{\mathbf q} + b_{-\mathbf{q}}^\dagger)$, where $V_{\mathbf q}$ is the electron-ripplon coupling, $\rho_{\mathbf q} = \sum_n e^{i \mathbf q \cdot \mathbf r_n}$ is the electron density operator, and $b_{\mathbf q}$ ($b_{\mathbf q}^\dagger$) annihilates (creates) a ripplon. We treat the electron-ripplon coupling to lowest nonvanishing order, while retaining the coupling to the ac drive nonperturbatively through the finite oscillation amplitude $\mathbf a$.

The first-harmonic frictional force can be written, in the low-temperature single-ripplon limit, as
\begin{align}
    \label{eq:force_general}
    \mathbf F(t)
    = -\frac{\cos(\omega t)}{\hbar n_s A}
    \sum_{\mathbf q}\sum_{m=1}^{\infty}
    \frac{m\mathbf q}{\alpha_{\mathbf{q}}}\,|V_\mathbf q|^2J_m^2(\alpha_{\mathbf{q}})\nonumber\\
    \times \frac{\bar n(\omega_q)+1}{\bar n(m\omega)+1}
    \mathcal S(\mathbf{q},m\omega-\omega_q),
\end{align}
where $\alpha_{\mathbf{q}}=\mathbf q\cdot\mathbf a$, $\omega_q$ is the ripplon frequency, $n_s$ is the electron surface density, $A$ is area of the microchannel, $\bar n(\omega)$ is the Bose-Einstein distribution, and $J_m(\alpha_{\mathbf{q}})$ are Bessel functions of the first kind. The function $\mathcal S(\mathbf q,\Omega)=\int_{-\infty}^{\infty}dt e^{i\Omega t}\langle{\rho_{q}(t) \rho_{-q}(0)\rangle}_0$ is the dynamic structure factor of the undriven Wigner crystal. In the electronic crystalline phase, the elastic component of the structure factor is strongly peaked at the Wigner crystal reciprocal-lattice vectors $\mathbf G$, i.e.
the ripplon emission is Bragg selected, and the sum over $m$ represents drive-induced sidebands generated by the periodic motion. 

In the weak-field limit, where $\alpha_{\mathbf{G}} \ll 1$ for the relevant reciprocal-lattice vectors, the first sideband dominates. Using $J_1(\alpha_{\mathbf G})\simeq \alpha_{\mathbf G}/2$, and neglecting higher $m$ terms, the zero-temperature frictional force reduces to
\begin{equation}
    \label{eq:force_zero_T}
    \mathbf F(t)
    \approx
    -\frac{n_sA\cos\omega t}{8\pi\hbar}
    \sum_{\mathbf G}
    \alpha_{\mathbf G}\mathbf G
    |\widetilde V_G|^2
    \delta(\omega-\omega_G),
\end{equation}
where $\widetilde V_G=V_G\exp(-G^2\lambda^2/4)$ includes the Debye-Waller suppression due to the zero-point fluctuations of the electrons with a mean-square  displacement $\lambda^2$. Eq.~\eqref{eq:force_zero_T} is the ac analogue of the Bragg-Cherenkov condition in the sense that the first harmonic of the driven density couples resonantly to ripplons whose frequency satisfies $\omega=\omega_G$. 

Despite the similarity, the force in Eq.~\eqref{eq:force_general} should not be identified directly with the quasi-static Bragg-Cherenkov effect of Ref.~\cite{Dykman97}, which is recovered only in the low-frequency regime where the motion during one cycle can be treated locally as having a fixed velocity. In contrast, the present high-frequency regime is intrinsically ac, i.e. the resonant condition is imposed by the drive frequency and its sidebands.  The characteristic ripplon frequency associated with the first reciprocal-lattice shell is of order $\omega_{G_1}/2\pi\sim 10\,{\rm MHz}$. Therefore, nonlinear features observed at substantially higher drive frequencies cannot, in general, be accounted for by the first reciprocal lattice shell alone. For a triangular Wigner crystal, the reciprocal-lattice vectors are $ \mathbf G_{hk}=h\mathbf b_1+k\mathbf b_2$, with magnitude $G_{hk}=G_1 \gamma_{h,k}$, where $\gamma_{h,k}=\sqrt{h^2-hk+k^2}$ and $h,k$ are integers. Using the ripplons dispersion relation, the Bragg frequencies scale as $\omega_{hk} = \omega_{G_1} \gamma_{h,k}^{3/2}$. Higher reciprocal-lattice shells therefore provide a natural kinematic route for extending the Bragg-Cherenkov condition into the tens-to-hundreds of MHz range. For example, the shell with $\gamma_{h,k}^2=3$ gives $\omega_{hk}/2\pi\simeq20-25\,{\rm MHz}$ if $\omega_{G_1}/2\pi\simeq10\,{\rm MHz}$, while frequencies near $150\,{\rm MHz}$ correspond to shells with $\gamma_{h,k}^2\simeq 37$. We note that this elastic-Bragg estimate identifies the reciprocal-lattice channels that are kinematically available, but the measured amplitude depends on the spectral weight carried by those channels. In particular, inelastic components of the Wigner-solid structure factor, coupling to Wigner-crystal phonons, finite crystalline correlation length, and dynamical pinning may broaden the resonances and redistribute spectral weight among different ripplon-emission channels.

In conclusion, we have performed high-frequency measurements to investigate the nonlinear transport response of a confined Wigner crystal on the surface of superfluid helium. We observe a nonlinear dynamical pinning of the crystal consistent with a frictional back-action arising from the coherent Cherenkov emission of ripplon overtones produced by higher-order Bragg vectors. Furthermore we observe a breakdown of this high-frequency pinned state, which we attribute to a non-equilibrium melting of the electron crystal. Future studies using even higher frequency driving fields are needed to learn the ultimate fate of the Wigner solid as its motional frequency approaches that of short wavelength plasmons (phonons) of the electronic solid, a regime in which quantum fluctuations will dominate the transport response.



We are grateful to M.I.~Dykman for valuable discussions and pointing us in the direction of this research. Additionally we acknowledge C.~Undershute and D.G.~Rees for valuable discussions. We also thank R.~Loloee and D.~Edmunds for technical assistance and B.~Bi for device fabrication advice and use of the W.M.~Keck Microfabrication Facility at MSU. The Michigan State University portion of this work was supported by the National Science Foundation via grant number DMR-2410650 and the Cowen Family Endowment at MSU. JP also acknowledges support from the Gordon and Betty Moore Foundation under Grant DOI~10.37807/GBMF13719.

\clearpage

\appendix

\section{Appendix A: Compensation line approach for high-frequency transport of electrons on helium in microchannel devices}

\begin{figure}
    \centering
    \includegraphics[width = 8.5cm]{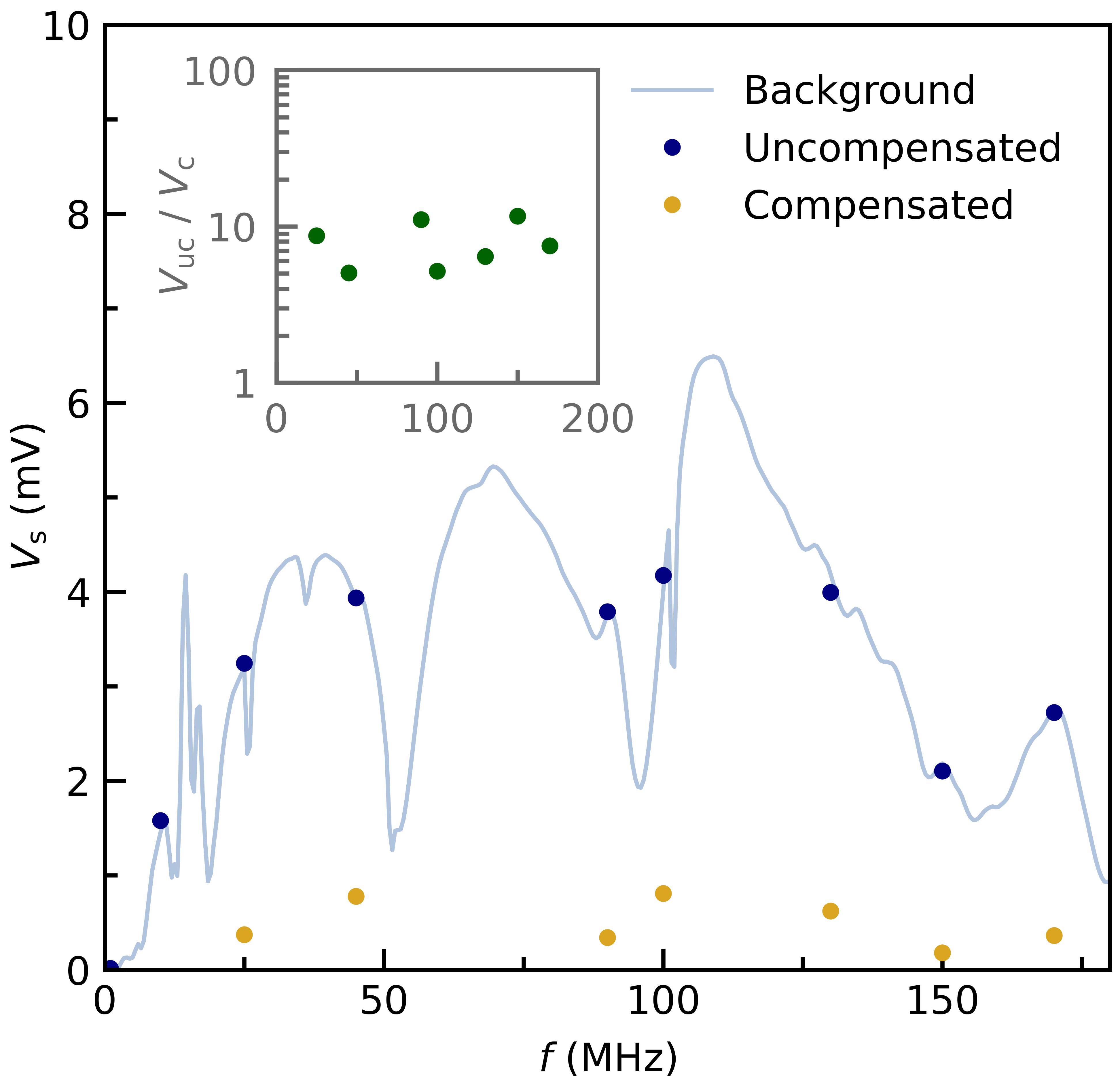}
    \caption{The experimentally measured parasitic background for $V_\text{ac} = 0.2~\text{V}$ both with (yellow) and without (blue) compensation for the parasitic background signal. The dots are associated with the transport measurements described in the main text while the solid line corresponds to a characterization of the entire background signal when the channel is devoid of electrons. Inset: Ratio of the uncompensated-to-compensated background signal, $V_{\textrm{UC}}/V_{\textrm{C}}$, as a function of drive frequency, showing an approximately tenfold decrease in the parasitic background.}
    \label{figA}
\end{figure}

With increasing drive frequency, transport experiments of electrons on helium in microchannel devices suffer from a parasitic background signal across the devices, which becomes more problematic with increasing frequency. A key contributing factor to this background signal is a capacitive cross-talk between the reservoir used to drive transport and the reservoir used to readout transport. In this work we implemented a ``compensation line'' approach in which we use destructive interference to attenuate the parasitic background. To achieve this reduction in background, we place a variable attenuator and a  variable phase-shifter in parallel to the microchannel device outside of the dilution refrigerator between the $V_\text{ac}$ input and the $V_\text{s}$ output (Fig.~\ref{fig3}a).

To perform the compensation line calibration, the central microchannel electrode is biased such that no electrons populate the microchannel ($V_\mathrm{ch}=0.3\ \mathrm{V}$) and the phase and amplitude of the compensation signal are tuned to destructively interfere with the remaining background signal.  Operationally, this tuning is accomplished by minimizing the overall signal $V_\text{s}$ as observed on an oscilloscope. In this fashion, the desired transport signal arising from from the dynamics of electrons in the central channel is enhanced relatively to the parasitic background.

Fig.~\ref{figA} presents a characterization of the performance of this compensation scheme as a function of measurement frequency. The solid line in Fig.~\ref{figA} is the measured parasitic background when the central channel contains no electrons. The blue dots are the values of frequency at which we performed the high-frequency transport measurements described in the main text. At each point we adjusted the variable attenuator and phase shifter to minimize the unwanted background. The yellow dots represent the background compensated signal after this procedure was performed. As shown in the inset, we find that this interferometric compensation scheme successfully reduced the parasitic background signal by approximately one order of magnitude.


\end{document}